# Security Issues in Vehicular Ad Hoc Networks (VANET): a survey

Yousef Al-Raba'nah, Ghassan Samara*

Department of Computer Science, Zarqa University, Zarqa, Jordan.
**Correspondence Address:** *Ghassan Samara, Department of Computer Science, Zarqa University, Zarqa, Jordan.
_______________________________________________________________________________________

**Abstract**
Vehicular Ad Hoc Networks (VANET) is a technology that has been recently emerged, and bring a lot of interests. VANET can be used to improve road safety, reduce road traffic, serve interests of its users, and provide emergency services. The security is one of the most important issues in VANET, it is considered a critical point in the development of robust VANET systems. In this paper, a diverse types of security challenges and requirements of VANET will be discussed, and a set of possible solutions for VANET security problems and attacks will be presented and analyzed. Also this paper will propose a new protocol that is called the reply protocol, this protocol aims to protect VANET against several attacks.

## Introduction

Rapid developments in wireless technologies provide possibilities to use these technologies in improvement of the driving environment, intending to allow road safety, infotainment and efficient transportation. Deaths are increasing dramatically in the world, and a significant proportion of these deaths lies on roads, around 1.2 million people are killed on roads yearly worldwide, and more than 50 million are injured. These numbers will increase by about 60% in next few years if no actions are taken[1], all of that in addition to other harms such as waste of time that is caused by traffic jams.

Vehicular Ad Hoc Networks (VANET) is a wireless networks, where vehicles are connected to each others, and can connect with internet. VANET is a special group of a Mobile Ad Hoc Networks (MANET), where nodes move freely, this means that there are no constraints on its movement. Each node will stay connected when it changes its location, as consequence VANETs have a highly dynamic topology. Nodes are communicating with each other in single hop or multi hop. Each node in VANET is either vehicle or Road Side Unit (RSU).

Communications in VANET are divided into two categories: Vehicle to Vehicle (V2V) communication, and Vehicle to Infrastructure (V2I) communication. In V2V, vehicle can communicate with other vehicles, involves sending and receiving messages to or from other vehicles. V2I takes place when vehicles communicate with RSU. These communications enable different applications to improve road safety and efficient transportation [4].

VANET is vulnerable to several attacks, because the nature of its open access medium. The security is the most important issue in VANET. To immune the members of VANET from attacks such as denial of service attack, many securing methods are proposed.

This paper investigates and analyzes the security challenges that are facing VANET and causing many problems, and presents comprehensive information about various VANET security requirements, challenges, possible attacks, and possible solutions that have been proposed.

The rest of this paper is organized as following: in section II we described VANET structure and how they are working, section III we list different attacks that threaten VANET and classified it, in section IV we identified the VANET challenges, like availability and mobility, section V discusses the security requirements that have to be met to enable secure system, in section VI we discussed different solutions that are proposed to solve the VANET problems, section VII discusses the proposed solutions, and section VIII concludes the paper.

## How VANET works

As mentioned above, nodes are forming VANET, and the number of these nodes are too large, currently there are more than 800 million vehicles in the world[2]. The communications between nodes done through using radio signals, range of these signals can reach up to 1 KM. Communications between nodes that have distance exceeds the signal range demand messages to hop across multiple nodes. Routing is done by a RSU, RSU plays as a router between vehicles. However, the following figure shows the VANET structure[3] and [15].

In order to connect vehicles with RSU using radio signals, each vehicle must be equipped with an On Board Unit (OBN). Tamper Proof Device (TPD) is a device that holds all vehicle secrets such as driver identity, speed, and position[3].

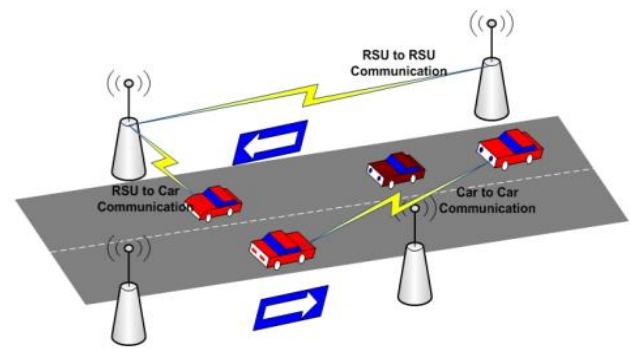

**Fig. 1: VANET structure [3]**

## Attacks and threats

The attacks against VANET affect its behaviour, to deal with these attacks, many researchers are classified these attacks. Researchers in [3], [6], [13], and [9] presented different classifications for attacks. Researchers in [6] classified attacks as attacks that pose a threat to availability, attacks that pose a threat to authenticity and attacks that pose a threat to driver confidentiality, and miscellaneous. While researchers in [13] classified attackers into three classes: Insider vs. Outsider, Malicious vs. Rational, and Active vs. Passive. Different classification proposed by researchers in [9], they classified attacks as Network Attack (NA), Timing Attack (TA), Monitoring Attack (MA), Social Attack (SA) and Application Attack (AP). Another classification suggested in [3], here researchers classified attackers into three classes: Selfish Driver, Malicious Attacker, and Pranksters. Each class describes different types of attacks. However, in this section we discuss and analyze some of these attacks.

### A. Denial of Service attack

Denial Of Service (DOS) attacks aim to make the network unavailable to its legitimate vehicles. Such attacks try to breakdown the network, and prevent sending and receiving of messages through the network to other vehicles. This happened when the attacker jams the network communication medium channel, or makes exclusive control of a vehicle's resources. In VANET, this attack is considered one of the most dangerous attacks, as the vehicles can't access the network and passing information messages to other vehicles[3], [5], [6], and [7].

**B. Sybil Attack**
This attack includes sending multiple copies of messages to other vehicles, and each message contain a fabricated identity, i.e. an attacker appears to other vehicles as hundreds of vehicles with different ids, telling them there is jam ahead and enforce them to take another route[3], [5], and [6].

**C. Alteration Attack**
Such attack happens when the attacker modifies or changes an existing data. This attack includes either delaying the transmitting of information messages, repeating previously transmitted messages, or modifying the original content of message and data transmitted[3], and [7].

**D. Message Suppression Attack**
Messages are sent and received as packets, an attacker can select some of these packets and dropping it. These packets may contain critical information. The attacker can keep these packets for later using. For example, an attacker might drop the jams alerts it receives, preventing it from being transmitted to the other vehicles, these alerts could help in selecting another path to destination, the vehicles that are not informed, enforce to wait in traffic. The dropped packets may be used again later to obtain the benefits. The main objective of the attacker is to deny the jams and collisions information from reaching authorities[3], [6], and [7].

**E. Identity Disclosure**
An attacker obtains the Identity (ID) of the target node and its current location to track them. This is achieved by sending malicious code to the adjacent of the target node to get the required data, as result the privacy of the target node will be disclosed. Rental companies use the tracked data to keep track of the movement of their vehicles[5], and [6].

**F. Spamming**
Spamming attacks aim to consume the network bandwidth and increase the transmission latency. The users are not interested in such messages, like advertisement messages[6].

**VANET Challenges**
  **A. Mobility**
  The vehicles in VANET have a highly dynamic topology, because it move freely, and during its movement it connect through the way with many vehicles that may never be faced before. The vehicles stay connected only for too short time, then the connection is lost as each vehicle moves toward its direction, this make the securing of VANET difficult to be achieved[3].

  **B. Volatility**
The connectivity among nodes in VANET can be highly fleeting, and live only for a limited period of time. As mentioned above, vehicles during movements can connect to other vehicles, so this connection will not stay for long time as the vehicles move freely, and change its movements directions. VANET lacks the relatively long-lived context, but contacting users device to a hot spot will demand long life password, something like this is impractical for securing vehicular communication[3], and [13].

  **C. Network Scale**
Current number of vehicles around the world exceeds 800 million, and this number are increased rapidly, which making the network scalable is difficult, and another problem will arise because the absence of a global authority existence that governs the standards for this network[8].

### D. Bootstrap
VANET is a promising technology, and requires that each vehicle equipped with Dedicated Short Range Communications (DSRC) radios, but until this moment, there are small number of vehicles satisfying this condition, so in developing applications for VANET, we must take into account that the communications will be limited to a few number of vehicles[3].

**Security Requirements in VANET**
In order to have a secure VANET, a number of security requirements must be satisfied, some of these requirements are needed to a network as a whole, and some of them are specific to VANET only. However, these requirements must be considered when designing vehicular network to hinder the attacks against VANET.

### A. Authentication
Means that to be able to send and receive messages through the network, VANET nodes must be authenticated. Authentication includes the process of verifying of the sender identity, and determine if he has the rights to communicate through the network[13].

### B. Availability
Availability needs high connectivity and bandwidth. the network must be available when it is needed, and sometimes it must have fast response time for specific applications, any delaying even if it takes milliseconds will make the message meaningless[6].

### C. Privacy
The personal and private information of drivers and vehicles must not be available to unauthorized access, i.e. immune of private information to be observed from unauthorized access.

### D. Integrity
In order, to trust messages contents, all messages that are sent and received through network must be protected against alteration attacks.

### E. Non-Repudiation
When sending a message, the attacker may wish to deny that sending, to avoid its responsibility. Non-repudiation enables identifying the attackers and prevents them from disavowal their crimes. This is achieved because all information are recorded and stored on TPD, so any authorized official side can retrieve this data[3].

**Security solutions**
To provide secure VANET, many researchers introduced a set of solutions to solve different security problems in VANET, researchers in [10], and [13] proposed the using of Vehicular Public Key Infrastructure (VPKI), here every node sends a safety message, it signs that message with its private key, and attaches it with Certificate Authority (CA). The receiver party of the message, will get the public key of the sending party by using the certificate, and check the signature of that sender, using its certified public key, but this solution requires that the CA public key be known by the receiver party.

Researchers [11] proposed using of group of signature, but this is not good solution, as a lot of overhead can be raised when using this solution. The idea behind this approach is that, there is a public key for a group of vehicles, and each vehicle has a session key, but when a new vehicle enters the group domain, the group public key as well as the vehicle session key of the vehicles in that domain must be altered and transmitted, also making the group static is difficult, as the vehicles have high mobility and continuously change their topology, as result the keys will be changed and transmitted frequently.

Another solution is proposed by [12], researchers suggested a new protocol for message checking, this protocol involves checking the Certificate Validity (CV) of the sender, the receiver of the message checks the CV of the message sender, the result of checking has three cases: in the first case, the receiver will consider the message if the sender has a valid certificate, second case occurs when the sender has invalid certificate, in this case the receiver will not regard the message, in the third case, the sender has not CV at all, the receiver will inform the RSU with the sender and check the received message, if it is correct the RSU will issue CV for the sender, otherwise it will issue invalid certificate and record vehicle's identity into the Certificate Revocation List (CRL). However, the next figure shows how the protocol works [12].

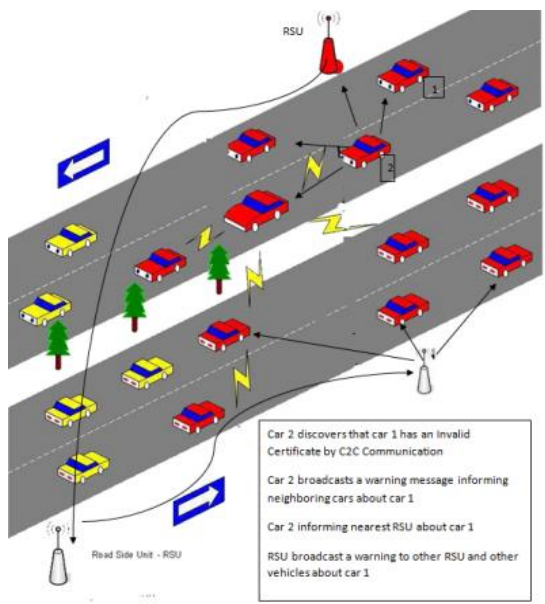

**Fig. 2: Message checking protocol[12]**

Other researchers proposed some solutions for different attacks. For example, to avoid DOF attacks, researchers in [8] suggested that switching between different channels or even communication technologies (e.g., DSRC, UTRA-Time Division Duplex (ULTRA-TDD), or even Bluetooth for very short ranges), if they are available, when one of them (typically DSRC) is brought down, while researchers in [13] said that frequency hopping do not completely solve the problem. The use of multiple radio transceivers, operating in disjoint frequency bands, can be a feasible approach.

To protect vehicular network against Sybil attacks, researchers in [14] proposed a solution, this solution involves using on road radar, where each vehicle can see surrounding vehicles and receive reports of their GPS coordinates. By comparing what is seen to what has been heard, a vehicle can corroborate the real position of neighbours and isolate malicious vehicles.

**Proposed solutions**
  **Reply Protocol**
In this protocol, when a vehicle sends a message to another vehicle, the receiver vehicle informs the nearest RSU, and requests it to check the correctness of the message, the RSU will communicate with the responsible RSU (which reside at the location of the sender vehicle), the latter RSU checks the message correctness by asking a random vehicle. Afterwards, the result will be distributed to all vehicles within the range of that RSU, and these vehicles will pass these messages through the way.

Another solution is proposed to protect vehicular networks against message suppression attack, the packets that are sent, attached with a time, this time indicating the initial time of packets sending, that time determines the age of these packets, if the attacker selects some of these packets and drops them, and tries to retransmit it again later, such packets become out of order, and other vehicles will not respond to it.

**Conclusion**
To make the development of VANET systems worth the effort, different security requirements and conditions must be satisfied. This paper has provided, discussed, and analyzed the VANET challenges, requirements, attacks, and its solutions. We have shown that the using of VPKI is giving the optimal solution. Also, we proposed a new protocol called the reply protocol, and we proposed another solution for message suppression attack. In the future work, we intend to suggest new solutions and protocols that can be enhance the VANET security and simulate these solutions.